\documentclass[aps,preprint,showpacs,superscriptaddress,groupedaddress,nofootinbib]{revtex4-1}  

\usepackage{color}
\usepackage{graphicx}  
\usepackage[colorlinks = true,
            linkcolor = blue,
            urlcolor  = blue,
            citecolor = blue,
            anchorcolor = blue]{hyperref}
\usepackage{multirow}
\usepackage{ulem}
\usepackage{amssymb}   
\usepackage{amsmath}
\usepackage{xspace}

\renewcommand{\emph}[1]{{\it #1}}
\newcommand{\mq}{\texttt{METAQ}\xspace}
\newcommand{\mpijm}{\texttt{mpi\_jm}\xspace}
\newcommand{\bash}{\texttt{bash}\xspace}
\newcommand{\thereadme}{\href{https://github.com/evanberkowitz/metaq/blob/master/README.md}{the \texttt{README}}\xspace}

\newcommand{\llnl}{
	Physics Division,
	Lawrence Livermore National Laboratory,
	Livermore, CA 94550, USA
	}
    
\newcommand{\fzj}{
	Institut f\"ur Kernphysik and Institute for Advanced Simulation,
	Forschungszentrum J\"ulich,
	52425 J\"ulich, Germany
	}

\begin{document}

\title{METAQ: Bundle Supercomputing Tasks}

\author{Evan~Berkowitz}
\email{e.berkowitz@fz-juelich.de}
\affiliation{\llnl}
\affiliation{\fzj}

\date{\today}

\begin{abstract}
We describe a light-weight system of \texttt{bash} scripts for efficiently bundling supercomputing tasks into large jobs, so that one can take advantage of incentives or discounts for requesting large allocations.
The software can backfill computational tasks, avoiding wasted cycles, and can streamline collaboration between different users.
It is simple to use, functioning similarly to batch systems like \texttt{PBS}, \texttt{MOAB}, and \texttt{SLURM}.
\end{abstract}

\maketitle

\section{Introduction}

High performance computing (HPC) is often performed on scarce, shared computing resources.  To ensure computers are being used to their full capacity, administrators often incentivize large workloads that are not possible on smaller systems.
For example, at the National Energy Research Scientific Computing Center (NERSC), jobs that take more than 683 nodes on Edison receive a 40\% discount on the consumed resources, while jobs on Titan at the Oak Ridge Leadership Computing Facility (OLCF) are artificially aged to ensure shorter wait times in the queue, and hence higher throughput and a shorter wall time to results, if they exceed 3750 nodes\cite{NERSC.charge,OLCF.charge}.

Some HPC tasks are done at supercomputing centers because they require \emph{capability}---that is, they require non-commodity hardware, such as fast communication interconnects---and some because they require \emph{capacity}---that is, there are many tasks to perform.  Measurements in lattice QCD (LQCD) often require a mix:  solves for quark propagators may require 8 or 32 high-performance nodes (or more), but do not typically require simultaneous resources on the scale of a full leadership-class machine.
However, the statistical nature of LQCD means that many solves are needed to complete a calculation.

The nontrivial advantages of running large jobs and the abundance of small or medium-sized computational tasks naturally leads to bundling---grouping many tasks into one job.  Throughout this discussion, by \emph{job} we mean an allocation provided by the batch scheduler, while a \emph{task} is an HPC step the user wishes to execute.

Because some tasks take longer than others and some nodes perform better than others, na\"ive bundling---simply starting many tasks and waiting for their completion before beginning new ones---leads to the potential for a lot of wasted computational cycles, offsetting any advantage or discount.  One solution, often adopted for similar reasons by batch schedulers, is to \emph{backfill}---to find smaller or shorter tasks that can be executed in the idle time left by a task's early completion.
Batch schedulers backfill entire jobs.  It would be useful for a user to be able to backfill the computational tasks themselves.
\mq, the software we describe, makes backfilling tasks easy.

\begin{figure}[t]
    \centering
        \includegraphics[width=0.45\textwidth]{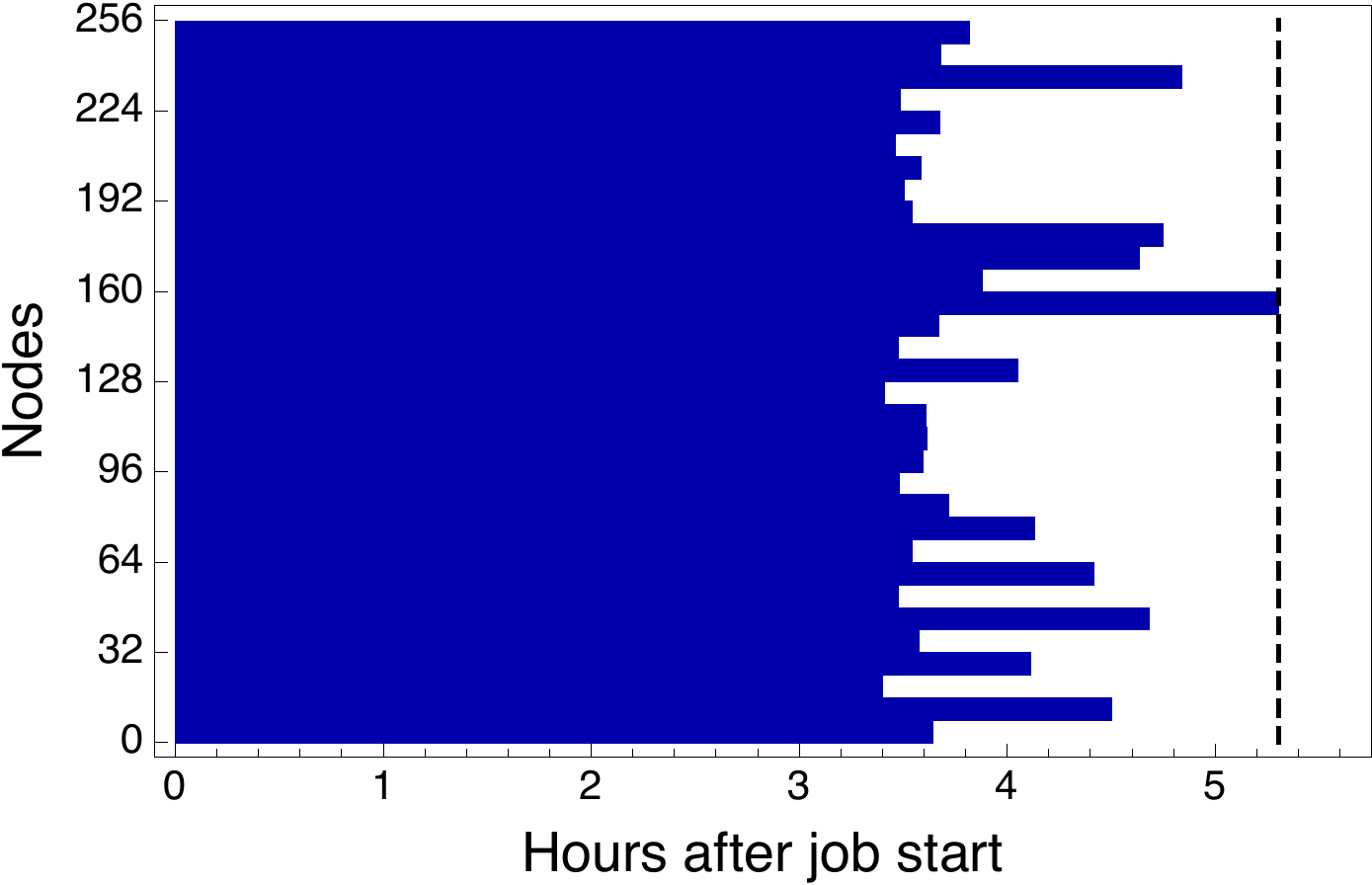}
        \includegraphics[width=0.45\textwidth]{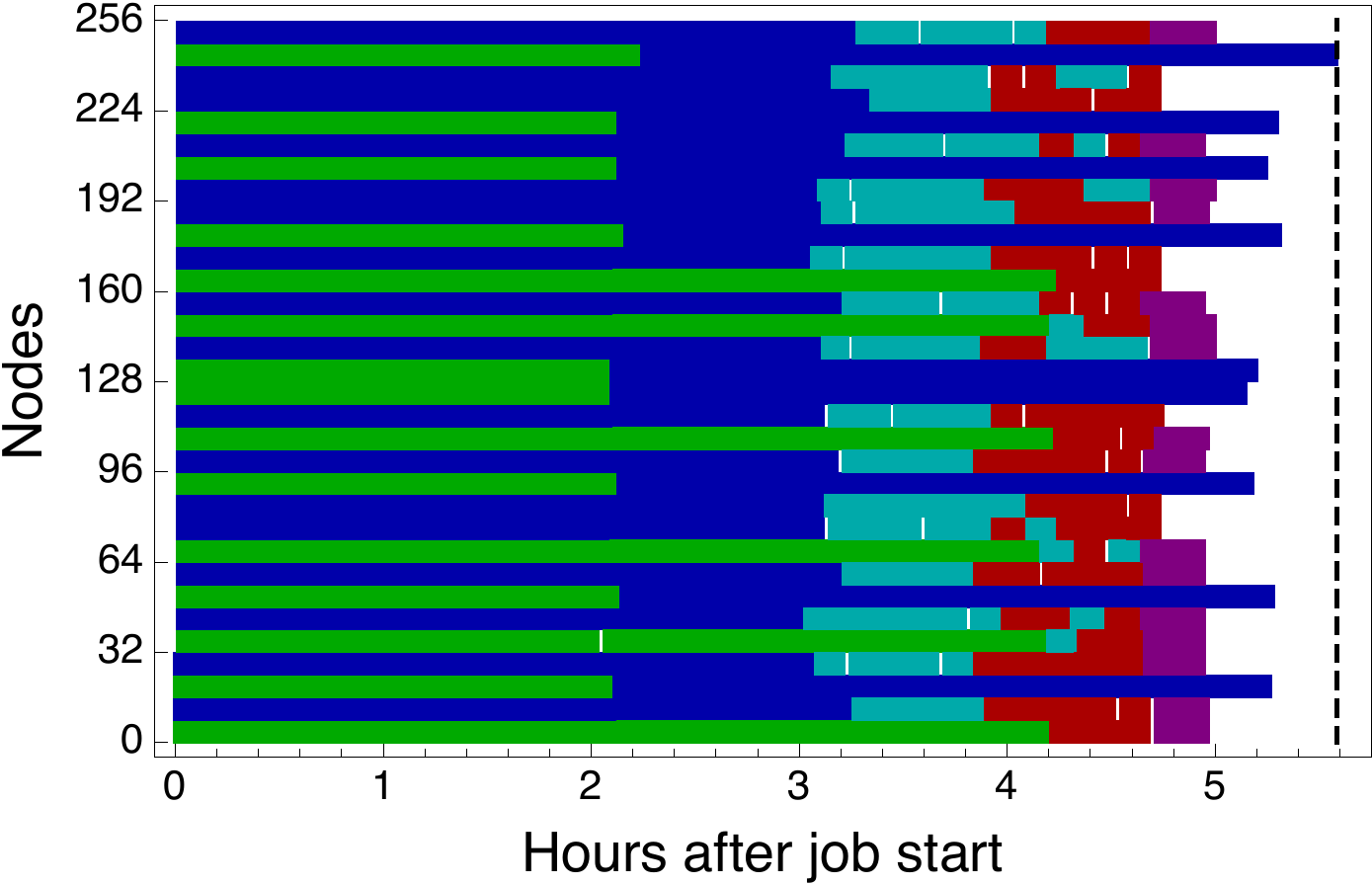}
    \linespread{1}\caption{
Two representative timelines of example jobs run on Titan.  
Different colors correspond to computations of different types, while white indicates nodes sitting idle.  The jobs terminated at time indicated by the vertical dashed line.
In the left panel, a substantial amount of compute time ($\sim28\%$) is wasted by a na\"ively-bundled group of 32 computationally-similar 8-node tasks because of the spread in run times.
In the right panel a mix different kinds of of 8-node tasks were launched at the job's start, and 8-node and 32-node tasks were intermixed and backfilled automatically by \mq, reducing the wasted time to $\sim10\%$.  Had the user (the author) given more accurate time estimates for red or purple tasks, there would have been even less idle time.
}
    \label{fig:backfill}
\end{figure}
\textbf{}

Additionally, coordinating high performace computations across many users and machines can pose a nontrivial problem.  Different computing environments might use different batch schedulers, incentivize different job sizes or wallclock times, or restrict the number of concurrent jobs any user can run.  All these can reduce the throughput and slow down the progress toward a completed computation.
It is useful, therefore, to separate the description of the computational tasks from the description of the jobs.

Here, we describe \mq (pronounced ``meta-queue''), a small, freely-licensed and freely-available\cite{berkowitz.metaq}, lightweight suite of \texttt{bash} scripts that alleviate many of these issues and that should be easily useable by anyone already familiar with the most popular HPC batch schedulers (eg. \texttt{SLURM}\cite{SLURM}, \texttt{MOAB}/\texttt{TORQUE}\cite{MOAB,TORQUE}, \texttt{PBS}\cite{PBS}).
\mq bundles task at run time, and allows you to easily backfill your jobs.
Overall, for our use cases, na\"ive bundling of similar tasks frequently resulted in $\sim$25\% wasted time (see the left panel of Fig.~\ref{fig:backfill}) simply because different nodes exhibited different performance or some problem instances proved harder than others, while using \mq we achieved closer to $\sim$5\% (right panel) idle waste.  This is computationally equivalent to a software speedup of about 25\%, as $\frac{1-0.05}{1-0.25} \simeq 1.27$.

\mq sits between the batch scheduler and the user's tasks.  
Users create scripts with markup that describes the resources and initialize any environment variables needed to perform the computing task, just as they would for any batch scheduler.
But rather than submit those scripts directly to the batch scheduler, users place them in specific folders.  
A separate script that specifies the size of the requested allocation and initializes \mq settings gets submitted to the batch scheduler.  We call scripts that get submitted to the scheduler \emph{job scripts} and scripts that describe the HPC work \emph{task scripts}.  
As mentioned earlier, a \emph{job} is an allocation provided by the batch scheduler, while a \emph{task} is a requested computation.

Factoring the description of work into jobs and tasks unlocks some additional nice features:
\begin{enumerate}
    \item Many users may create task scripts, and jobs can bundle the tasks created by different users together.
    \item Many users can submit job scripts to the batch scheduler, each attacking the same set of tasks.
    \item Tasks for different projects may be bundled together simply.
\end{enumerate}
On the other hand, this factorization is not very advantageous for tasks without easy parallelism, such as the Markov-chain generation of LQCD gauge configurations.

Separating the submission and running of jobs from the specification of the computational workload has the nice advantage that many users can submit task scripts to the inspected folders, and every user's tasks can be picked up by any user's job.  This allows collaborators to streamline the running of jobs---rather than assign one person one portion of the work, every collaborator can easily submit job scripts to the batch scheduler and build up priority on the HPC system, while the generation of tasks can be scripted and handled by fewer users.

Moreover, it enables you to \emph{change} what tasks will be done before a job starts without cancelling the job and losing its accumulated priority.
In fact, you can submit jobs to build priority before deciding what tasks to run.
On machines where large jobs might begin many days after submission, the ability to alter the computational workload can be extremely useful and save substantial human time.

In the remainder of this work, we describe \mq in more detail---borrowing heavily from \thereadme---and discuss where this kind of workflow management is heading in the future.

\section{Job Scripts}

Job scripts are \bash scripts that are submitted to the batch scheduler.
They perform any set-up, allow the user to specify options, and then attack the outstanding tasks.
This makes them relatively easy to read, write, and modify.

When writing a job script, the user is required to specify some configuration variables and may optionally specify others.  Configuration variables always begin with `\texttt{METAQ\_}'.
A detailed enumeration of all the options \mq understands is provided in \thereadme and basic template job scripts are provided for PBS, SLURM, and the mixed MOAB/SLURM environment at LLNL, as \texttt{x/run.pbs.sh}, \texttt{x/run.slurm.sh}, and \texttt{x/run.llnl.sh} respectively.

Here we review just the required options.
\begin{enumerate}
    \item \texttt{METAQ} should specify the full path to the \mq folder (that is, the root directory of the cloned \texttt{git} repo).
    \item \texttt{METAQ\_JOB\_ID} can be any string, as long as you ensure that it is unique.  In most cases one simply uses the batch scheduler's job id (eg. \texttt{\$SLURM\_JOB\_ID}).
    \item \texttt{METAQ\_NODES} should usually describe how many nodes you requested from the batch scheduler.\footnote{See Sec.~\ref{gpu.vs.node} for discussion of an instance where lying to \mq was advantageous.}
    \item \texttt{METAQ\_RUN\_TIME} indicates how long the job will run.  It can be an integer, which is interpreted as seconds, or it can follow the format \texttt{[[HH:]MM:]SS} which is understood by the command \texttt{x/seconds}.  
    
So, you do not have to specify canonically-formatted times. For example, you can specify \texttt{90:00} or \texttt{5400} instead of \texttt{1:30:00}.  However, \texttt{1:30} will be parsed as a minute and a half.

Depending on your batch scheduler, it may be possible to have \mq detect the allocated wallclock time, in which case it may be possible to specify this setting automatically.  Otherwise, when writing a new job script, one may have to specify the time twice---once as an instruction to the batch scheduler, and once as a \mq setting.  A mismatch between these two times might create a scenario where time is wasted.

    \item \texttt{METAQ\_MACHINE} can be any string, but is canonically the name of the actual computer.  It is currently only used to help organize the working directory.  
\end{enumerate}

Of the optional configuration variables, here it is worth discussing \texttt{METAQ\_TASK\_FOLDERS} which is a \bash array of priority-ordered absolute paths of folders where you want \mq to look for tasks.  This defaults to \texttt{(\$METAQ/priority \$METAQ/todo)}, but can be easily changed or configured for more advanced usage.

Job scripts can in principle reside anywhere on disk, but conventionally are placed in the \texttt{x} directory.  Moreover, the default \texttt{git} settings for the \mq repo ignore files that start with \texttt{x/q\_}, so that you can create job scripts without triggering changes to the core \mq software.

Job scripts can do task-independent preliminary work.  For example, you can have a job script resubmit itself (for a chained workflow) or query a database to populate task folders with computational tasks.  But, you should keep in mind that any time spent in this kind of setup step is time where the compute nodes are idle.

Finally, once the configuration is set, one simply invokes \texttt{source \${METAQ}/x/launch.sh} to attack the queue of task scripts.  Anything after this invocation is not guaranteed to run.  For example, if the wallclock time runs out, the job will be cancelled while still in the \mq main loop.

\section{Task Scripts}

Tasks are the high performance computing steps.
It is these task scripts that contain the command that tells the batch scheduler to put work onto the compute nodes (for example, in a \texttt{SLURM} setting these task scripts contain \texttt{srun} commands).

Task scripts must be directly executable---you must include the correct shebang and ensure it can be executed successfully on the computer's service nodes.
Task scripts need not be \bash. The structure of a task script is only very loosely constrained.
One should not rely on inheriting environment variables---for clarity they should be set in the task script itself, as different tasks might need different environments.

You tell \mq about the task by incorporating flags that \mq understands into the task script. A bare-bones \bash task script that needs 2 nodes for 5 minutes looks schematically like 
\linespread{1}
\begin{verbatim}
#!/bin/bash
#METAQ NODES 2
#METAQ MIN_WC_TIME 5:00
#METAQ PROJECT metaq.example.1

# do setup:
load modules
set environment variables
etc

echo "working hard..."
[your usual call to srun, aprun, etc. which takes 2 nodes in this example]
echo "finished!"
\end{verbatim}
where the command to launch your executable on the HPC nodes is exactly what would appear in a normal run script.
\mq ignores any line that does not have `\texttt{\#METAQ}' in it, so one may construct task scripts that are also well-formed for submission directly to the batch scheduler.  \mq flags need not come at the top of a task script.

Flags are always \texttt{CAPITALIZED}, preceeded by `\texttt{\#METAQ}', and live on their own line.  One may put flags inside the comments of other languages, as long as there is no whitespace between the beginning of the line and the `\texttt{\#METAQ}' annotation (so that the value of the flag is always in the third \texttt{awk}-style field).  \mq only reads the first instance of a flag---if you put the same flag on multiple lines, all the others will be ignored.

The flags that \mq understands are
\begin{enumerate}
    \item \texttt{\#METAQ NODES N} --- This task requires \texttt{N} nodes, where \texttt{N} is an integer.
    \item \texttt{\#METAQ GPUS  G} --- This task requires \texttt{G} GPUs, where \texttt{G} is an integer.
    \item \texttt{\#METAQ MIN\_WC\_TIME [[HH:]MM:]SS} --- The job must have \texttt{HH:MM:SS} \underline{w}all\underline{c}lock time remaining to start this task.  An accurate estimate helps to avoid having work that fails due to interruption and helps to avoid wasting time.
    
    You may specify times in the format understood by the command \texttt{x/seconds}, which converts a passed string into a number of seconds, as mentioned in the discussion of the \texttt{METAQ\_RUN\_TIME} setting above.
    \item \texttt{\#METAQ LOG   /absolute/path/to/log/file} --- Write the running log of this task to the file specified.
    \item \texttt{\#METAQ PROJECT  some.string.you.want.for.accounting.purposes} --- \mq does not use the task's project for anything substantive. However, it does log the project string to \texttt{jobs/\${METAQ\_JOB\_ID}/resources}. This is convenient if you have many comingled projects (or parts of projects) in the same \mq, so that you can account for how the computing time was spent.

For easy sorting and organization, it often makes sense to have the string prefixed in order of generality (most specific detail last, like reverse domain name notation).
\end{enumerate}

\subsection*{What is meant by a ``\texttt{GPU}'' and what is meant by a ``\texttt{NODE}''?}
\label{gpu.vs.node}

\mq does not actually interface with the compute nodes of a machine.
Instead, it keeps track of what resources are used and what resources are free locally on a service node.
This opens an ambiguity and potential source of confusion when describing the available hardware to \mq.

On many machines there are only a few GPUs and many CPUs on a single physical node. It may be possible to overcommit, meaning that you can run one binary on a few CPUs and the GPUs and another binary on the remaining CPUs.
If you cannot overcommit (eg. on Titan at OLCF), then you can forget about \texttt{GPU}s altogether and just worry about \texttt{NODE}s, and think of them as the physical nodes on the machine.

If you can overcommit, then potentially what you should mean by \texttt{GPU} is (one GPU and one controlling CPU), while a \texttt{NODE} is (\#CPUS per node - \#GPUs per node). For example, on LLNL's GPU machine Surface each node has 2 GPUs and 16 CPUs\cite{LLNL.surface}. So, from the point of view of \mq there are 2 \texttt{GPU}s and 1 14-CPU \texttt{NODE}s per hardware node.  Alternately, it might be that you expect any GPU-capable job to use both GPUs, in which case you can think of 1 \texttt{GPU} and 1 14-CPU \texttt{NODE} per hardware node.  Or, you might expect a GPU-capable job to require both GPUs and 8 CPUs, in which case you can think of 1 \texttt{GPU} and 1 8-CPU \texttt{NODE} per hardware node.  Whatever choice you make is fine---as long as you are consistent across all of your task scripts.\footnote{We found that the ability to overcommit successfully sometimes relies on the GPU-enabled jobs launching first.}

It may be that your binaries are smart, and can simultaneously make good use of all of the resources on a physical node.  In this case, you can forget about \texttt{GPU}s altogether.

This ambiguity can also be used advantageously.  Users can set up a task that could run on different machines that see the same \mq. However, currently \mq does not know how to read machine-dependent settings. In the example where we first understood this aspect of \mq, one machine had nodes with huge memory while the other didn't, meaning that on the first we could run on one physical node and on the other we needed four physical nodes. By writing a task script that branched based on the machine, setting \texttt{\#METAQ NODE 1}, and setting the initialization variable \texttt{METAQ\_NODES} to one-quarter of the number of physical nodes on the machine with less memory, we could circumvent this hardware requirement mismatch and use the two different machines to work on the same queue of tasks.

\section{Installation, Directory Structure, and Basic Usage}

A basic installation can be accomplished with a simple \texttt{git clone} of the repo into any directory that is readable and writeable from the supercomputing nodes.  A fresh clone has only one directory, which contains all the \mq scripts.

If multiple users will be collaborating on one set of tasks, one should set appropriate permissions.  It often makes sense to recursively apply sticky group permissions (\texttt{chmod -R g+rwXs}) to the whole repo.

Run \texttt{x/demo.sh} from the root directory.  It creates the other standard directories, and populates the \texttt{todo} directory with a small set of dummy tasks (where the ``hard work'' is simply \texttt{sleep}ing for a short, random amount of time).

That is all that is required for a functional installation.  If you want to take advantage of some of the other provided utilities (see Sec.~\ref{sec:util}), run \texttt{x/install} from the root directory.  You will be prompted for a choice of batch scheduler, which you should pick to match your computing environment.  Currently, \texttt{PBS} and \texttt{SLURM} are understood.  You can add support for another batch scheduler with minimal effort, as explained in greater detail in \thereadme.

By default, there are three folders where task scripts are placed by the user:  \texttt{priority} (which is inspected for tasks first), \texttt{todo} (inspected next), and \texttt{hold} (not inspected at all).  Once complete, the task scripts are moved to the \texttt{finished} folder.

We have found that it makes sense to organize tasks by their computational requirements.  For example, it might make sense to put all the 32-\texttt{NODE} 0-\texttt{GPU} tasks together, separate from 8-\texttt{NODE} 8-\texttt{GPU} tasks.
The particular organization is left to the user.

In each folder in \texttt{METAQ\_TASK\_FOLDERS}, \mq looks for a special file, \texttt{.metaq}.  It parses that file for the \texttt{\#METAQ} flags \texttt{NODES}, \texttt{GPUS} and \texttt{MIN\_WC\_TIME}, just as it would a task script, to short-circuit the need to check the requirements of every task script in that folder.
For example, if \mq knows it only has 4 \texttt{NODES} available and it is currently looking through a folder whose \texttt{.metaq} claims that the folder contains jobs that require 8 \texttt{NODE}s, it will skip to the next folder.
\mq does not \emph{enforce} the consistency of the folder's \texttt{.metaq} file and the task scripts that folder contains.
So, if misled, \mq might skip over a task that it could otherwise launch.
If the \texttt{.metaq} file is not there, \mq will loop over every file in the folder no matter what.

When a task gets picked up by \mq it is moved out of the queue and into a working directory, \texttt{working/\${METAQ\_MACHINE}/\${METAQ\_JOB\_ID}}.  The atomicity of disk operations\footnote{
Task scripts are moved with \texttt{mv}, which ensures disk atomicity as long as the computer's operating system does not crash and as long as the files are moved on the same filesystem.
} thereby ensures that two different jobs that are running simultaneously cannot execute the same task---which would result in work duplication (and, in the worst case, potentially corrupt results).

\mq logs the output of every task to the log optionally specified by the \texttt{\#METAQ LOG} flag in the task script, as well as \texttt{jobs/\${METAQ\_JOB\_ID}/log/\${TASK\_FILE\_NAME}.log}.  

\mq keeps track of the available and consumed resources (\texttt{NODES} and \texttt{GPUS}) in the \texttt{jobs/\${METAQ\_JOB\_ID}/resources} file, which is actually symbolically linked to a timestamped file (to avoid confusion in the event a job gets interrupted and restarted)\footnote{This was a very difficult issue to track down, because it happened only rarely and only at certain computing facilities.}.  This file is relatively verbose for ease of human interpretation, but the first column on every line is always a number and the form is relatively rigid, so it can be parsed by machine with ease.

\section{The \mq Main Loop}

When the batch scheduler provides an allocation of computing resources, \mq configures according to user-specified variables in the job script and then proceeds to look for tasks in the folders specified by \texttt{METAQ\_TASK\_FOLDERS}.

\mq understands it only has a finite number of nodes (and optionally and separately, GPUs) and a limited wallclock time to undertake computational tasks.
It will not try to launch tasks that require more than the resources currently free or that will require more than the remaining wallclock time (so as to avoid interrupted work).
\mq attempts to find and launch tasks that fit into the currently-available resources, automatically back-filling tasks.

The main loop of \mq can be described by the pseudocode
\linespread{1}\begin{verbatim}
loop over all possible remaining tasks until there are none:
    Check if you currently have enough resources (nodes, GPUs,
        clock time, etc.) to perform the task
    If so,
        move it to the working directory, 
        deduct those resources from what's available, 
        and launch it!
    Otherwise, skip it---
        But, if it is impossible, don't count this job as "remaining".
            Some examples of impossibility:
                The task needs more nodes than are allocated to this job.
                The remaining clock time isn't enough to complete the task.
\end{verbatim}

Frequently called \bash functions are stored in \texttt{x/metaq\_lib.sh}, and the main execution logic is in \texttt{x/launch.sh}.

\section{Additional Utilities}
\label{sec:util}

In this section, we discuss some of the additional small utilities provided with \mq.

\begin{itemize}
    \item \texttt{x/report} - This utility helps break down by project the expended resources (\texttt{NODE}-hours and \texttt{GPU}-hours), by parsing the \texttt{resources} files in the \texttt{jobs} directory and reporting a few basic metrics.
    
    \item \texttt{x/running} - This utility takes an argument, which is the name of the \texttt{METAQ\_MACHINE} you are currently concerned with.  If \mq is batch-scheduler aware, \texttt{x/report} can report which \mq jobs in the working directory are still running and which have been abandoned.
    
    \item \texttt{x/reset} - This utility takes an argument, which is the name of the \texttt{METAQ\_MACHINE} you are on.  If \mq is batch-scheduler aware, \texttt{x/reset} will compare the \texttt{\$HOSTNAME} to \texttt{\^\$argument.*\$}.  If there is a match, it will place tasks from abandoned jobs in the \texttt{priority} folder.  Otherwise, it will issue a warning and prompt for user confirmation.  If you use multiple machines that cannot see a common batch scheduler to work on the same set of tasks, this is a risky utility, because you might move currently-running tasks out of their appropriate \texttt{working} folder.
    
    \item \texttt{x/status} - This utility can give a quick-glance summary of the outstanding tasks in the \texttt{priority}, \texttt{todo}, and \texttt{hold} directories (and their sub-directories), producing a count based on the tasks' \texttt{PROJECT} flags.

\end{itemize}

\section{Future Development}

While not perfect, \mq is largely feature-complete, and very little substantial future development is foreseen.  

\mq was designed so as to change the user experience on HPC resources as little as possible.  Users still write scripts for every task and put them in the queue.  This has the advantage that essentially nothing needs to be changed, and executables need not be recompiled.

The success of \mq inspired work on \mpijm\cite{mpijm}, a \texttt{C++} replacement for \mq that lives at the MPI level so that it stresses the service nodes of machines much more lightly and only relies on a single \texttt{aprun} (or \texttt{srun}, or equivalent).  We hope a first production-quality release of \mpijm will be finished sooner rather than later, at which point development of \mq will likely cease entirely, though I am happy to review and accept pull requests. There may still be some situations where \mq may be preferable---since \mpijm will require recompiling against it as a library (and adding a handful of lines of code to your executable).  The advantage of moving to this sort of model is the ability to more finely manage the computing resources while still being able to write separate executables that need not do their own management of enormous partitions, and to be able to run CPU-only and GPU-capable executables simultaneously on a given node without the batch scheduler's blessing.  This manager-worker model may prove useful as the next generation of supercomputers come online and we head towards the exascale.

\mpijm can be compiled against any MPI library that provides \texttt{MPI\_Comm\_spawn}\cite{MPI}.
We have also integrated a \texttt{Python} interface, so task management logic and task description can make use of high-level language constructs and can easily \texttt{import} \texttt{Python} modules.
Ultimately, we expect an extremely fruitful connection between \mpijm and standard databases, which can further streamline and coordinate nontrivial computations.

\section{Known Bugs, Flaws, Complaints, and Missing Features}

\mq launches each task script as its own process on a service node and keeps it alive until completion.  So, each task requires its own \texttt{aprun} (or equivalent) and a few processes of overhead on the service nodes.

\mq has run well, although not perfectly in every situation.  On Titan, our run model stressed the service nodes too heavily and we ultimately crashed the machine, although we were in compliance with the user guide (which was subsequently amended to limit the number of processes a job can launch).  This is another motivation for the development of \mpijm---it will cost only one \texttt{aprun} (or equivalent) and put all of the task management on a compute node.  Morever, it is written in \texttt{C++}, which is much more performant and low-level than \bash.  Ultimately, we ran on smaller partitions on Titan to stay within the new guidelines, forefeiting the aging bonus.  Nevertheless, we found \mq helped reduce the brain cycles spent on running jobs there.

\mq logs its own behavior and that of all tasks.  From time to time many gigabytes of logs can build up and a manual purge of the \texttt{log} and \texttt{jobs} directory is needed.  From the \mq root directory this can be accomplished with, for example, a simple invocation of
\begin{verbatim}
    find ./jobs/ ! -newermt YYYY-MM-DD -delete -print
\end{verbatim}
for the \texttt{jobs} directory, where \texttt{YYYY-MM-DD} is an ISO 8601-style date.  For safety users should avoid deleting any file that any active job is writing.

Another shortcoming is the lack of a task submission command, like \texttt{msub}, \texttt{qsub}, or \texttt{sbatch} which could automatically sort tasks into different folders based on their resource requirements.  As it stands, users put scripts directly into the folders, and \mq does not enforce any consistency between the \texttt{.metaq} folder description files and the other task scripts.

Bugs may be filed and \href{https://github.com/evanberkowitz/metaq/issues}{issues may be opened on \texttt{github}}.

\section*{Acknowledgements}

\mq grew out of finding lattice QCD calculations increasingly difficult to coordinate, and is a ground-up rewrite of purpose-written run scripts I originally crafted with Thorsten Kurth that performed very inflexible na\"ive bundling.  The separation of job submission from task description was emphasized as a useful trick by Chris Schroeder when I began working on lattice QCD at LLNL.

This work was performed under the auspices of the U.S. Department of Energy by Lawrence Livermore National Laboratory (LLNL) under Contract DE-AC52-07NA27344 and \mq was tested on \texttt{aztec}, \texttt{cab}, \texttt{surface}, and \texttt{vulcan} at LLNL through the Multiprogrammatic and Institutional Computing and Grand Challenge programs. \mq was also tested and used in production on \texttt{titan} at Oak Ridge Leadership Computing Facility at the Oak Ridge National Laboratory, which is supported by the Office of Science of the U.S. Department of Energy under Contract No. DE-AC05-00OR22725, and \texttt{edison} and \texttt{cori} at NERSC, the National Energy Research Scientific Computing Center, a DOE Office of Science User Facility supported by the Office of Science of the U.S. Department of Energy under Contract No. DE-AC02-05CH11231.

This work was supported in part by the Office of Science, Department of Energy, Office of Advanced Scientific Computing Research through the CalLat SciDAC3 grant under Award Number KB0301052.

The first large-scale use of \mq was lattice QCD calculations by the CalLat collaboration, and enabled us to expedite our neutrinoless double beta decay calculation\cite{2016:0vbb_proceedings} through our production pipeline with minimal effort and concern without interrupting other ongoing calculations\cite{Berkowitz:2017opd}.  
\mq allowed us to efficiently consume our 65M Titan-node hour INCITE allocation in approximately 5 months using tasks of 32 nodes or less, bundled efficiently in larger jobs of 256 to 512 nodes.  I am indebted to Jason Chang, Amy Nicholson, Enrico Rinaldi and Andr\'{e} Walker-Loud who helped stress-test \mq, proved its worth in production, and provided valuable feedback and bug reports.  I am excited to complete \mpijm with Ken McElvain, Thorsten Kurth, and Andr\'{e} Walker-Loud.

\bibliographystyle{apsrev} 
\bibliography{metaq}

\begin{thebibliography}{12}
\expandafter\ifx\csname natexlab\endcsname\relax\def\natexlab#1{#1}\fi
\expandafter\ifx\csname bibnamefont\endcsname\relax
  \def\bibnamefont#1{#1}\fi
\expandafter\ifx\csname bibfnamefont\endcsname\relax
  \def\bibfnamefont#1{#1}\fi
\expandafter\ifx\csname citenamefont\endcsname\relax
  \def\citenamefont#1{#1}\fi
\expandafter\ifx\csname url\endcsname\relax
  \def\url#1{\texttt{#1}}\fi
\expandafter\ifx\csname urlprefix\endcsname\relax\def\urlprefix{URL }\fi
\providecommand{\bibinfo}[2]{#2}
\providecommand{\eprint}[2][]{\url{#2}}

\bibitem[{NER()}]{NERSC.charge}
\emph{\bibinfo{title}{How usage is charged}},
  \bibinfo{howpublished}{\url{http://www.nersc.gov/users/accounts/user-accounts/how-usage-is-charged/}},
  \bibinfo{note}{accessed: 2017-01-23}.

\bibitem[{OLC()}]{OLCF.charge}
\emph{\bibinfo{title}{Titan scheduling policy}},
  \bibinfo{howpublished}{\url{https://www.olcf.ornl.gov/kb_articles/titan-scheduling-policy/}},
  \bibinfo{note}{accessed: 2017-01-23}.

\bibitem[{\citenamefont{Berkowitz}(2016)}]{berkowitz.metaq}
\bibinfo{author}{\bibfnamefont{E.}~\bibnamefont{Berkowitz}},
  \emph{\bibinfo{title}{\texttt{METAQ}}},
  \bibinfo{howpublished}{\url{https://github.com/evanberkowitz/metaq}}
  (\bibinfo{year}{2016}).

\bibitem[{\citenamefont{SchedMD}(2006-2017)}]{SLURM}
\bibinfo{author}{\bibnamefont{SchedMD}}, \emph{\bibinfo{title}{{SLURM: Simple
  Linux Utility for Resource Management}}},
  \bibinfo{howpublished}{\url{https://slurm.schedmd.com/}}
  (\bibinfo{year}{2006-2017}).

\bibitem[{\citenamefont{{Adaptive Computing}}(2004)}]{MOAB}
\bibinfo{author}{\bibnamefont{{Adaptive Computing}}},
  \emph{\bibinfo{title}{Moab}},
  \bibinfo{howpublished}{\url{http://www.adaptivecomputing.com/products/hpc-products/moab-hpc-basic-edition/}}
  (\bibinfo{year}{2004}).

\bibitem[{\citenamefont{{Adaptive Computing}}(2003-2017)}]{TORQUE}
\bibinfo{author}{\bibnamefont{{Adaptive Computing}}},
  \emph{\bibinfo{title}{{TORQUE: Terascale Open-source Resource and QUEue
  Manager}}},
  \bibinfo{howpublished}{\url{http://www.adaptivecomputing.com/products/open-source/torque/}}
  (\bibinfo{year}{2003-2017}).

\bibitem[{\citenamefont{{Altair Engineering}}(1991-2017)}]{PBS}
\bibinfo{author}{\bibnamefont{{Altair Engineering}}},
  \emph{\bibinfo{title}{{PBS: Portable Batch System}}},
  \bibinfo{howpublished}{\url{https://github.com/pbspro/pbspro}}
  (\bibinfo{year}{1991-2017}).

\bibitem[{LLN()}]{LLNL.surface}
\emph{\bibinfo{title}{Surface}},
  \bibinfo{howpublished}{\url{https://hpc.llnl.gov/hardware/platforms/surface}},
  \bibinfo{note}{accessed: 2017-02-13}.

\bibitem[{\citenamefont{{Ken McElvain, Evan Berkowitz, Thorsten Kurth,
  Andr\'{e} Walker-Loud}}(2017)}]{mpijm}
\bibinfo{author}{\bibnamefont{{Ken McElvain, Evan Berkowitz, Thorsten Kurth,
  Andr\'{e} Walker-Loud}}}, \emph{\bibinfo{title}{\texttt{mpi\_jm}}},
  \bibinfo{howpublished}{in preparation} (\bibinfo{year}{2017}).

\bibitem[{MPI(2015)}]{MPI}
\emph{\bibinfo{title}{\texttt{MPI}: A message-passing interface standard,
  version 3.1}},
  \bibinfo{howpublished}{\url{http://mpi-forum.org/docs/mpi-3.1/mpi31-report.pdf}}
  (\bibinfo{year}{2015}), \bibinfo{note}{accessed: 2017-02-14}.

\bibitem[{\citenamefont{{Amy Nicholson, Evan Berkowitz, Chia Cheng Chang, M. A.
  Clark, B\'{a}lint Jo\'{o}, Thorsten Kurth, Enrico Rinaldi, Brian Tiburzi,
  Pavlos Vranas, Andr\'{e} Walker-Loud}}(2016)}]{2016:0vbb_proceedings}
\bibinfo{author}{\bibnamefont{{Amy Nicholson, Evan Berkowitz, Chia Cheng Chang,
  M. A. Clark, B\'{a}lint Jo\'{o}, Thorsten Kurth, Enrico Rinaldi, Brian
  Tiburzi, Pavlos Vranas, Andr\'{e} Walker-Loud}}},
  \bibinfo{journal}{PoS(LATTICE 2016)}  (\bibinfo{year}{2016}),
  \eprint{\href{https://arxiv.org/abs/1608.04793}{hep-lat/1608.04793}},
  \urlprefix\url{https://arxiv.org/abs/1608.04793}.

\bibitem[{\citenamefont{{Evan Berkowitz, Chris Bouchard, Chia Cheng Chang, M.A.
  Clark, B\'{a}lint Jo\'{o}, Thorsten Kurth, Christopher Monahan, Amy
  Nicholson, Kostas Orginos, Enrico Rinaldi, Pavlos Vranas, and Andr\'{e}
  Walker-Loud}}(2017)}]{Berkowitz:2017opd}
\bibinfo{author}{\bibnamefont{{Evan Berkowitz, Chris Bouchard, Chia Cheng
  Chang, M.A. Clark, B\'{a}lint Jo\'{o}, Thorsten Kurth, Christopher Monahan,
  Amy Nicholson, Kostas Orginos, Enrico Rinaldi, Pavlos Vranas, and Andr\'{e}
  Walker-Loud}}} (\bibinfo{year}{2017}), \eprint{1701.07559}.

\end{thebibliography}

\end{document}